\documentclass[psfig,apalike]{article}
\usepackage{url}
\usepackage{epsfig}
\usepackage{apalike}
\usepackage{english}

\newcommand{\comment}[1]{}

\begin{document}

\newcommand{\argmax}{{\rm argmax}}
\newcommand{\argmin}{{\rm argmin}}

\newcommand{\R}{{\sf I}\!{\sf R}}
\newcommand{\N}{{\sf I}\!{\sf N}}
\newcommand{\RSV}{{\rm RSV}}
\newcommand{\sRSV}{{\rm  RSV}}
\newcommand{\ERSV}{{\rm ERSV}}
\newcommand{\NRSV}{{\rm nRSV}}
\newcommand{\logit}{{\rm logit}}

\newcommand{\snake}{\sim}
\newcommand{\ndf}{{\rm ndf}}
\newcommand{\E}{{\rm E}}
\newcommand{\Var}{{\rm Var}}
\newcommand{\nff}{{\rm nff}}
\newcommand{\eff}{{\rm eff}}
\newcommand{\icf}{{\rm icf}}
\newcommand{\iecf}{{\rm iecf}}
\newcommand{\ecf}{{\rm ecf}}
\newcommand{\ff}{{\rm ff}}
\newcommand{\df}{{\rm df}}
\newcommand{\hpe}{{\rm hpe}}
\newcommand{\fap}{{\rm fap}}
\newcommand{\cf}{{\rm cf}}
\newcommand{\ncf}{{\rm ncf}}
\newcommand{\idf}{{\rm idf}}
\newcommand{\nidf}{{\rm nidf}}
\newcommand{\LBC}{{\rm LBC}}
\newcommand{\RC}{{\rm RC}}
\newcommand{\rc}{{\rm rc}}
\newcommand{\RCC}{{\rm RCC}}
\newcommand{\NAP}{{\rm nAP}}
\newcommand{\NL}{{\rm nL}}
\newcommand{\LL}{{\rm L}}
\newcommand{\AP}{{\rm AP}}
\newcommand{\n}{{\rm n}}
\newcommand{\nl}{{\rm nl}}
\newcommand{\XX}{{\rm {\cal X}}}
\newcommand{\SUB}{{\rm SUB}}
\newcommand{\INS}{{\rm INS}}
\newcommand{\DEL}{{\rm DEL}}

\newcommand{\rank}{\rho}
\newcommand{\nrank}{{\rm n\rho}}
\newcommand{\norm}{{\rm norm}}

\newenvironment{bitemize}
  {\begin{itemize} \setlength{\itemsep}{1cm}}%
  {\end{itemize}}
\newenvironment{benumerate}
  {\begin{enumerate} \setlength{\itemsep}{1cm}}%
  {\end{enumerate}}

 \newenvironment{nitemize}
  {\begin{itemize} \setlength{\itemsep}{0cm}}%
  {\end{itemize}}
\newenvironment{nenumerate}
  {\begin{enumerate} \setlength{\itemsep}{0cm}}%
  {\end{enumerate}}

\title{A New Ranking Principle for \\Multimedia Information Retrieval}
\author{Martin Wechsler, Peter Sch\"auble  \\
        Eurospider Information Technology AG, Z\"urich \\
        \url{www.eurospider.com} }
\date{}
\maketitle
\begin{abstract}
A theoretic framework for multimedia information retrieval is
introduced which guarantees optimal retrieval effectiveness.  In
particular, a Ranking Principle for Distributed Multimedia-Documents
(RPDM) is described together with an algorithm that satisfies this
principle.  Finally, the RPDM is shown to be a generalization of the
Probability Ranking principle (PRP) which guarantees optimal retrieval
effectiveness in the case of text document retrieval.  The PRP
justifies theoretically the relevance ranking adopted by modern search
engines.  In contrast to the classical PRP, the new RPDM takes into
account transmission and inspection time, and most importantly,
aspectual recall rather than simple recall.

\end{abstract}

\label{ch:u}

\section{Introduction}
\label{intro}

Multimedia Information Retrieval is becoming more and more feasible
because both speech and image recognition methods have been improved
significantly during the last years:  A wealth of new information
access techniques have been developed to find relevant information in
large multimedia data collections \cite{Schauble:97}. While some of
 these new techniques -- for example retrieval techniques for digitized
 speech documents \cite{Wechsler:98a} -- have been evaluated
 experimentally, they were hardly developed and
studied within a theoretic framework that guarantees optimal retrieval
effectiveness; in fact we lack such a framework that optimizes the
probability that a user finds the desired information in a large
multimedia document collection.  In this paper, a Ranking Principle
for Distributed Multimedia (RPDM) document collections is described
that serves as a theoretic framework like the Probability Ranking
Principle (PRP) by Robertson (1977) for centrally stored text
documents. The PRP states that \nocite{Robertson:77}{\em a retrieval system
performs optimally if the documents are ranked according to decreasing
probabilities of relevance}. He showed that optimal performance can be
expressed either in terms of precision and recall, or in terms of
costs associated with the retrieval of non-relevant documents and the
non-retrieval of relevant documents. We elaborate on the PRP in
Section~\ref{sc:i.rp.prp}.

Nowadays, large {\em distributed multimedia} document collections are
available. Local and global computer networks, for example the
World-Wide Web, allow quick access and transfer of documents
independent of their location. Further, new technologies enable
digital processing of non-text media, such as images, audio and video,
which collectively emboss the notion of {\em multimedia}.

\section{Criteria for Distributed Multimedia Documents}

For IR systems managing distributed multimedia documents, we believe
that the optimal document ranking problem has to be revised. Such IR
systems should allow for {\em additional criteria} other than solely
the probability of relevance when suggesting an inspection order for
documents with respect to a query. We identify the following two
additional criteria:
\begin{enumerate}
\setlength{\itemsep}{-0.15cm}
\item {\em Transmission time} of a document, the time needed to 
transport a document from the source location across the network to
the user.
\item {\em Inspection time} of a document, the time needed by the user to
inspect a document.
\end{enumerate}

We demonstrate the importance of these criteria by the following two
examples that are also illustrated in Figure~\ref{fig:i.rpm}:
\begin{figure}[htbp]
  \begin{center} 
    \psfig{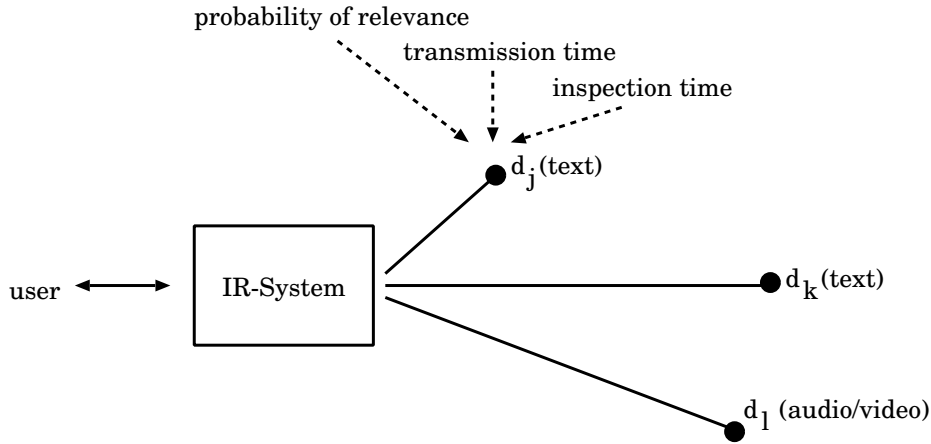}
    \caption{Retrieval of distributed multimedia documents and parameters affecting optimal ranking.}  
    \label{fig:i.rpm} 
  \end{center}
\end{figure}
(1) Assume that two documents $d_j$ (text) and $d_k$ (text) have an
equal probability of relevance to a given request. If $d_j$ is
geographically closer to the user than $d_k$, then obviously $d_j$
should be transmitted and presented before $d_k$, such that the user's
idle waiting time is minimized. (2) Assume that two documents $d_k$
and $d_l$ contain the same information but are from different media,
say $d_k$ is text and $d_l$ is audio or video. In this case the IR system
should favor the ranking of the text document first, since it is much
faster both to transmit and to inspect.

The transmission time is affected (1) by the geographical distance,
(2) by the network bandwidth, and (3) by the document's storage size,
which depends on the document's medium and length. The media text,
audio and video are orders of magnitude apart from each other with
regard to their associated data rates. Table~\ref{tab:i.dr} shows the
data rates for different (un)compressed formats of the three media
\cite[p.49,108]{Lu:96}:
\begin{table}[htbp]
  \begin{center}
    \leavevmode
    \begin{tabular}{|l|l|l|} \hline
     document medium & data format & data rate \\\hline\hline
     text     & ASCII & 0.018  kByte/s \\\hline
     audio & uncompressed CD-audio & 176.4 kByte/s\\\hline
     audio & compressed CD-audio (MPEG-audio) & 29.4 kByte/s \\\hline
     video & uncompressed VHS-video & 6750 kByte/s \\\hline
     video & compressed VHS-video (MPEG-1) & 187.5 kByte/s \\\hline
    \end{tabular}
    \caption{Data rates for different compressed and uncompressed media.}
    \label{tab:i.dr}
  \end{center}
\end{table}
The text data-rate is based on a speaking rate of 120 words per minute
and an average word-length of 9 letters. For audio and video,
sophisticated compression techniques such as MPEG \cite{Pan:95} make a
data reduction possible, however the differences between the media are
still considerable.

The inspection time is dependent on the medium and the length of the
document as well. Due to the time-synchronous nature of audio and
video, these media
require much more time for inspection compared with for example an
equivalent text document. Also, it has been found experimentally that
users require more time to extract requested information from a video
than from a text \cite{Sutcliffe:97}.

\section{Ranking Principle}
Before we study the optimal ranking problem in an IR system, we make
two general assumptions: First, we assume that the IR system is able
to return (query-dependent) {\em passages} rather than documents in
response to a query. Passages are motivated mainly by the fact that
entire audio and video documents require very much time for
transmission and inspection, and by the fact that a user would like to
listen only to relevant parts of a (maybe long) recording.
The second assumption is that
a query $q$ consists of $k$ aspects $q_i, i=1,...,k$. For example, the
query ``W.A. Mozart'' may ask for information about Mozart's
compositions but also about his life in society.

To study the optimal ranking problem for distributed multimedia
documents we assume that a user specifies a {\em total inspection
time} when he submits a query. We further assume that, after query
evaluation, the IR system has the following parameters available for
each passage: (1) the probability of relevance, which is estimated by
any retrieval method, (2) the transmission time, and (3) the
inspection time. 

\label{pg:rpdm}
\begin{center}
\fbox{\parbox{0.7\textwidth}{{\bf Ranking Principle for\\Distributed Multimedia Documents (RPDM)}\\
\\
An IR system should present passages of distributed multimedia
documents to a user query in such a way that
\begin{enumerate}
\setlength{\itemsep}{-0.1cm}
\item
The passages can be inspected within the user-specified total
inspection time,
\item
There is no user waiting time between the inspection of passages due
to their transmission,
\item The passages contain ``a maximum amount of relevant information
about various aspects of the query''.
\end{enumerate}
}}
\end{center}

More formally, we formulate the RPDM as an optimization problem. Given
are the
\begin{itemize}
\setlength{\itemsep}{-0.1cm}
\item total inspection time $T$ specified by the user,
\item the set $X$ of passages $x_j, j=1,...,|X|$ found to query $q$,
\item transmission time $t^t(x_j)$ for each passage $x_j$, 
\item inspection time $t^i(x_j)$ for each passage $x_j$,
\item probabilities of relevance $P(R | q_i,x_j)$ for each passage $x_j$ with regard to a query aspect $q_i$.
\end{itemize}
The problem for the IR system is to find a sequence $Y =
\langle y_1,...,y_n\rangle$ of passages $y_j \in X$ such that the following conditions are satisfied:
\begin{eqnarray}
  \label{eq:i.wut} 
  t^t(y_1) + \sum\limits^{n}_{j=1} t^i(y_j) \le T,  && \\
  \label{eq:i.nw} 
  t^t(y_j) \le t^t(y_1) + \sum\limits^{j-1}_{l=1} t^i(y_l) \,\,\, \forall j: 1\le j \le n,  && \\
  \label{eq:i.cy} 
  C(Y)  :=  \sum\limits_{i=1}^{k}
              \prod\limits_{y \in Y} (1-P(R | q_i,y)) \stackrel{!}{=} \min. &&
\end{eqnarray}
The first condition~(\ref{eq:i.wut}) requires that all selected
passages can be transmitted and inspected within the total inspection
time $T$. 

\begin{figure}[bp]
  \begin{center} 
    \psfig{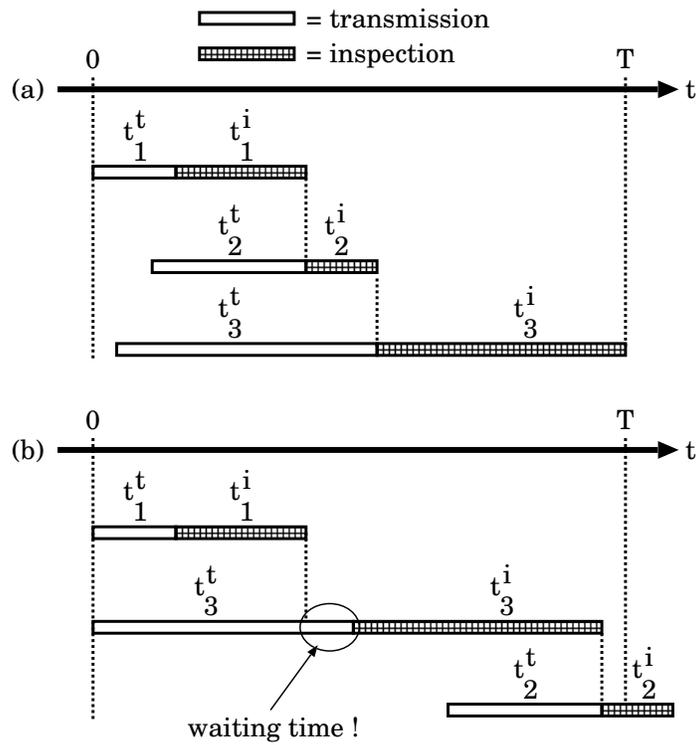}
    \caption{User inspection process for three passages. Future passages are transmitted in the background. (a) No user waiting time, (b) user waiting time between the first and second passage. Note also that in (b), the last passage cannot be inspected completely.}  
    \label{fig:inspproc} 
  \end{center}
\end{figure}

The second condition~(\ref{eq:i.nw}) assures that the next passage to be
inspected can be transmitted completely within the time used for the transmission of the first passage and the
inspection of all preceding passages. Here we assume that the IR
system is capable of requesting future passages in the background
while the user is inspecting. This eliminates waiting time due to the
transmission of passages. Figure~\ref{fig:inspproc} illustrates two
time scenarios of the user's inspection process, where
condition~(\ref{eq:i.nw}) is satisfied only in scenario (a).

Definition~\ref{eq:i.cy} is the main {\em cost function} which has to
be {\em minimized}. We justify the choice for the cost function as
follows: The term $(1-P(R | q_i,y))$ denotes the probability that
passage $y$ does {\em not} cover aspect $q_i$, and thus the product
denotes the probability that aspect $q_i$ is not at all covered in the
selected passages.  If we associate constant costs for each aspect not
covered, the cost function $C(Y)$ is proportional to the expected
costs for missing relevant aspects of the query. In other words, the
IR system should suggest passages that optimally cover the most
aspects to the given query. Our cost function is inversely proportional to
the aspectual recall used in TREC's interactive track \cite{Vorhees:99}. 
In the next section we present an
algorithm that solves this optimization problem.

\section{An Algorithm Satisfying the RPDM}
\label{sc:i.alg}

The simplest algorithm to the problem formulated in the previous
section is a {\em backtracking procedure} \cite{Nievergelt:77}\cite{Kreher:98}. The idea
of backtracking applied to our situation is to {\em enumerate all}
sequences of passages (i.e. possible solutions) while (1.) checking
both conditions~(\ref{eq:i.wut}) and (\ref{eq:i.nw}) for each sequence,
(2.) computing its cost function $C(Y)$, and (3.) retaining the
sequence that obtains minimal costs.

We show, however, that backtracking is not feasible for this problem
since the number of solutions to be considered is far too high. Let
$N=|X|$ be the total number of possible passages that have been found
to a given query, and assume that the algorithm selects $n$ passages.
\begin{figure}[htbp]
  \begin{center} 
\leavevmode 
\fbox{
\begin{minipage}{\textwidth}
\begin{tabbing}
\hspace{0.8cm}\=\hspace{0.4cm}\=\hspace{0.4cm}\=\hspace{0.4cm}\=\hspace{0.4cm}\= \kill
 \> \\
 \>  VAR\\
 \>\>$Y_{best}$; /* currently best solution */\\
 \>\>$C_{best}$; /* costs of best solution */\\
 \>\\
1 \>  PROCEDURE branch\_and\_bound( Sequence $Y$ ) \\
2 \>\>  VAR\\
3\>\>\> $C_Y$; /* costs for solution $Y$ */\\
4 \>\>\> $\LBC(Y_{x}^{+})$; /* lower bound costs for extensions of $\langle Y,x\rangle$ */\\
5 \>\>\> $E_Y$; /* set of possible passage extensions of $Y$ */ \\
6\\
7 \>\>  $C_Y := C(Y)$;\\
8 \>\>  IF $C_Y < C_{best}$\\
9 \>\>\>  $Y_{best} := Y; C_{best} := C_Y$  \\
10 \>\>  END;\\
11 \>\>  FOR ALL $x \in X \setminus {\rm set}(Y)$\\
12\>\>\>  IF {\em within\_user\_time}($\langle Y,x\rangle$) $\wedge$ {\em no\_waiting}($\langle Y,x\rangle$)\\
13\>\>\>\>  calculate $\LBC(Y_x^{+})$;\\
14\>\>\>\>  $E_Y$ := $E_Y \cup \{x\}$;\\
15\>\>\>  END  \\
16\>\>  END;  \\
17\>\>  LOOP\\
18\>\>\>  $x$ := \argmin$\{\LBC(Y^{+}_x) \, | \, x \in E_Y \wedge \LBC(Y^{+}_x) < C_{best}\}$\\
19\>\>\>  IF $x$ undefined \\
20\>\>\>\>  BREAK;\\
21\>\>\>  END\\
22\>\>\>  {\em branch\_and\_bound}($\langle Y,x\rangle$);\\
23\>\>  END; \\
24\>\>  RETURN \\
25\> END PROCEDURE;\\
 \>\\
 \>  /* INPUT: */\\
 \>  /* $X$: set of passages found to a request */\\
 \>  /* $t^t(x)$: transmission times  */\\
 \>  /* $t^i(x)$: inspection times  */\\
 \>  /* $P(R|q_i,x)$: probabilities of relevance */\\
 \>  /* $T$: total user inspection time  */\\
 \>  MAIN\\
 \>\>  $Y_{best} := \epsilon; C_{best} := \infty;$\\ 
 \>\>  {\em branch\_and\_bound}($\epsilon$);\\
 \>  END\\
 \>  /* OUTPUT: $Y_{best}$ */
\end{tabbing}
\end{minipage}
}
\caption{Recursive branch-and-bound algorithm satisfying the RPDM. }
\label{fig:bbalg}
\end{center}
\end{figure}
The number of solutions with $n$ passages can be derived by combining
$n$ out of $N$ passages and by permutating those sequences, which
results in
\begin{eqnarray}
  {N \choose n} \cdot n! & = & \frac{N!}{(N-n)!}  \nonumber
\end{eqnarray}
solutions. Since $n$ may vary from 1 to $N$, the {\em total number of solutions}
to be considered is
\begin{eqnarray}
  && \sum\limits_{n=1}^{N} \frac{N!}{(N-n)!}.  \nonumber
\end{eqnarray}
For $N=10$ this value is $\approx 10^7$ and for $N=100$ it is
$\approx 2\cdot10^{158}$. 

\begin{figure}[htbp]
  \begin{center}
  \psfig{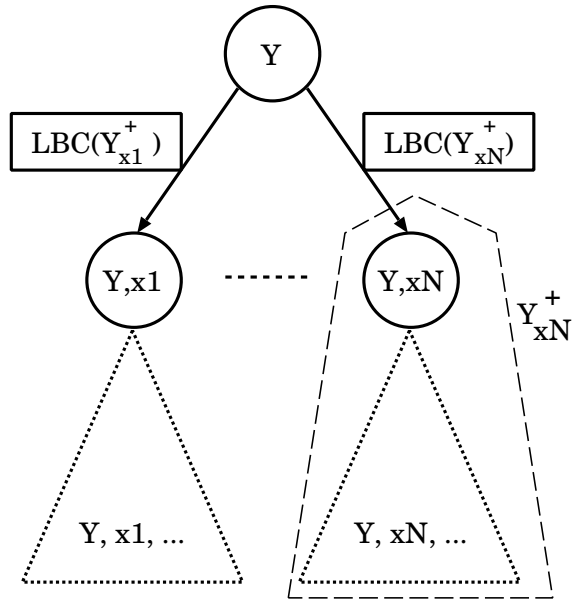}
  
\caption{Search tree of the branch-and-bound algorithm. Each node represents a
solution. Successor nodes are extensions of the solution $Y$. The
LBC-values are used either to select an extension as the next solution, or to discard an entire subtree.}

  \label{fig:i.bbfig} 
\end{center}
\end{figure}

To reduce the number of solutions to consider, we propose the classic
{\em branch-and-bound algorithm} (e.g. \cite[p.114--119]{Domschke:91}), which
is a variant of the backtracking algorithm. The idea of
branch-and-bound is (1) to enumerate solutions with lowest {\em
expected cost-values} first (branch), and (2) to discard entire sets
of solutions where costs can certainly not be further reduced
(bound). The pseudo-code of the algorithm is given in
Figure~\ref{fig:bbalg}. The algorithm is shown as a recursive procedure {\em
branch\_and\_bound} that calculates the cost function for a given
sequence $Y$ and tries to extend this sequence with an additional
passage at the end, before it is called recursively.  The main call of
this procedure is {\em branch\_and\_bound($\epsilon$)}, where
$\epsilon$ denotes the empty sequence. Figure~\ref{fig:i.bbfig}
illustrates the extension of the sequence $Y$ graphically. The circles
correspond to solutions and the arrows denote extensions with a single
passage.

In line~7 of the algorithm, the costs $C_Y$ of the current solution
are calculated according to the cost function
(Definition~\ref{eq:i.cy}). Then, the currently best solution
$Y_{best}$ is updated if necessary. The rest of the procedure
considers extensions of $Y$ with an additional passage $x$. We write
$\langle Y,x\rangle$ for a new sequence consisting of $Y$ and an
appended passage $x$. Also, we define $Y^{+}_x$ as the set of all
sequences that consist of the sequence $Y$, an appended passage $x$
and any number of further passages. In lines~11-16, for each extension
passage $x$ and each resulting set $Y^{+}_x$, a {\em lower bound of
costs} $\LBC(Y^{+}_x)$ is calculated (line~13, see next paragraph),
and the extension passage is retained in an extension set $E_Y$
(line~14). However, this is only performed if two conditions are met
(line~12): The procedure {\em within\_user\_time($Y,x$)} tests the
condition~(\ref{eq:i.wut}), and {\em no\_waiting($Y,x$)} tests the
condition~(\ref{eq:i.nw}). Finally, in lines~17-23 the extension with
the lowest LBC-value is pursued by a recursive call of {\em
branch\_and\_bound} (branch). In line~18, the best passage is selected
for extension, and simultaneously, extensions are only pursued if
their costs (represented by the lower bound costs) may become smaller
than the costs of the currently best solution $C_{best}$ (bound).

In the following, we derive a {\em lower bound of costs}
$\LBC(Y^{+}_x)$ for all solutions consisting of the solution $Y$
extended with a passage $x$ and any number of further passages:
According to the cost function (Definition~\ref{eq:i.cy}) we write for
the costs of solution $Y$
\begin{eqnarray}
  C(Y) & := & \underbrace{\prod\limits_{y \in Y} (1-P(R | q_1,y))}_{P(q_1)} 
              + \ldots + \underbrace{\prod\limits_{y \in Y} (1-P(R | q_k,y))}_{P(q_k)} \nonumber
\end{eqnarray}
where $P(q_i)$ abbreviates a product term. If we extend the solution
$Y$ with a passage $x$, it holds that
\begin{eqnarray}
  \label{eq:i.cyx} 
  C(\langle Y,x\rangle) &:= & P(q_1)\cdot (1-P(R | q_1,x)) 
               + \ldots + P(q_k)\cdot  (1-P(R | q_k,x)) \nonumber \\
       & \ge  & C(Y) \cdot \min_{i=1,...,k}\{1-P(R | q_i,x)\}. 
\end{eqnarray}
Now we consider any solution $Y' \in Y^{+}_{x}$, which starts with the
sequence $Y$ followed by the passage $x$ and any number of further
passages. From Equation~\ref{eq:i.cyx} follows:
\begin{eqnarray}
  \label{eq:i.cy+} 
  C(Y') & \ge  & C(Y) \cdot \min_{i=1\ldots k}\{1-P(R | q_i,x)\} \cdot \nonumber \\
  && \underbrace{(\min\{1-P(R | q_i,x') \,|\, i=1,...,k \wedge x' \in X \setminus {\rm set}(\langle Y,x\rangle)\})^m}_{=:\,\,r(Y,x)},
\end{eqnarray}
where $r(Y,x)$ is an abbreviation and $m$ denotes the maximum number
of passages that may be added to the sequence $\langle
Y,x\rangle$. This number is limited by the remaining inspection time
and by the minimum inspection time of the passages not yet
selected. Let $t_{Y}$ be the time necessary to inspect all passages of
the solution $Y$. Then we can formulate an upper bound for $m$ by
\begin{eqnarray}
  m & \le  & \left\lfloor\frac
                {T - t_Y - t^i(x)}
                {\min\{t^i(x') \,|\, x' \in X \setminus {\rm set}(\langle Y,x\rangle)\}}\right\rfloor. \nonumber
\end{eqnarray}
Thus, the right part of Equation~\ref{eq:i.cy+} contains a lower bound
for the costs of a solution $Y' \in Y_{x}^{+}$ and therefore we can
define
\begin{eqnarray}
  \LBC(Y^{+}_{x}) & := & C(Y) \cdot \min_{i=1\ldots k}\{1-P(R | q_i,x)\} \cdot r(Y,x).\nonumber
\end{eqnarray}
The lower bound costs denote the minimum possible costs of any
solution extended from a solution $Y$ and the passage $x$. If they are
larger than the currently minimum costs, then the entire search
subtree $\langle Y,x\rangle$ (Figure~\ref{fig:i.bbfig}) may be
discarded.

The output of the branch-and-bound algorithm satisfies the RPDM as
stated on page~\pageref{pg:rpdm}. A crucial element of the
branch-and-bound algorithm is the derivation of a highest possible
lower bound for the costs of extended solutions. The higher the lower
bound is, the more efficient the algorithm is because the number of
solutions to consider is reduced. However, more accurate lower bounds
are usually more complex to compute. This results in a trade-off
between the effort to compute a lower bound and the effort to consider
more solutions.

In a practical situation it may be that the IR system does not have
enough time to run the branch-and-bound algorithm because the search
tree is still too large. For such cases we propose the use of
suboptimal stochastic optimization techniques (e.g. simulated
annealing \cite{Vidal:93} or genetic algorithms
\cite{Mitchell:96}). These approaches do not guarantee that the best
solution will be found, but they may find a sufficiently good solution
in a short time.

\section{The RPDM and Robertson's PRP}
\label{sc:i.rp.prp}

In this section we show that our ranking principle for distributed
multimedia collections is compatible to Robertson's {\em Probability
Ranking Principle} (PRP) which was developed in the context of
retrieval in rather short texts and bibliographic records
\cite{Robertson:77} .

We map our distributed multimedia model into the context of the PRP,
which originally was to retrieve bibliographic references or (rather
short) text documents. We make the following assumptions:
\begin{itemize}
\setlength{\itemsep}{-0.1cm}
\item 
We do not deal with passages but with documents because the (text)
documents are very short (less than 1 page).
\item
The transmission time of each document is neglected ($t^t(d_j)=0
\,\, \forall j$) because each document consists of only a few text words.
\item The inspection time $t^i(d_j)$ for each document is constant,
say $\Delta t$, because the short text documents are able to be
inspected at a glance.
\item
There is only one aspect in the query.
\end{itemize}

Again, we assume that the user specifies a total inspection time
$T$. With constant inspection times, it follows that the user inspects
$n := \lfloor\frac{T}{\Delta t}\rfloor$ documents. Let $Y$ be the
sequence of those $n$ documents. Since the transmission times are
neglected, the documents of $Y$ may be presented in any order.

The cost function (Definition~\ref{eq:i.cy}) is simplified due to the
presence of only one aspect:
\begin{eqnarray}
  \label{eq:i.cyprp} 
  C(Y) & := & \prod\limits_{d_j \in Y} (1-P(R | q,d_j)). 
\end{eqnarray}

{\bf Proposition:} If the solution $Y$ contains the {\em top} $n$
documents with respect to $P(R|q,d_j), d_j \in Y$ (and thus conforms
to the PRP), it follows that its costs $C(Y)$ are minimal.

{\bf Proof (by contradiction):} Assume that the solution $Y$ contains
the top $n$ documents, and that its costs are {\em not} minimal. We
can choose any document $d \in Y$ and replace it by any other document
$d' \notin Y$ from the rest of the collection. According to
Definition~\ref{eq:i.cyprp} the costs of this new set $Y'$ are
\begin{eqnarray}
  C(Y') & := & \frac{1-P(R | q,d')} {1-P(R | q,d)} \cdot
               \prod\limits_{d_j \in Y} (1-P(R| q,d_j)) \nonumber \\
        & = & \underbrace{\frac{1-P(R | q,d')} {1-P(R | q,d)}}_{=: f} \cdot C(Y) . \nonumber
\end{eqnarray}
Since $P(R|q,d) \ge P(R|q,d')$, it follows $f\ge 1$ and thus
$C(Y')\ge C(Y)$ or $C(Y)$ is minimal. q.e.d.

The statement made in the proposition conforms to the Probability
Ranking Principle under the assumption that the number of documents to
inspect is known a priori. Thus we have shown that the context of the
Probability Ranking Principle is a special case of our RPDM, and that
the PRP is satisfied if the RPDM is satisfied.

\section{Conclusions}

We have introduced a new {\em ranking principle for
distributed multimedia documents} (RPDM). This principle describes the
optimal ranking of distributed multimedia documents by taking into
account not only the documents' probabilities of relevance, but also
their transmission and inspection times. We have formulated the
ranking principle as an optimization problem, and we have presented an
algorithm with which the principle can be satisfied. Lastly, we have
shown that the RPDM is a generalization of Robertson's probability
ranking principle.


\bibliographystyle{xapalike}
\bibliography{my_bibl,totbibl,eubib}

\begin{thebibliography}{}

\bibitem[Domschke \& Drexl, 1991]{Domschke:91}
Domschke, W., \& Drexl, A. (1991).
\newblock {\em {Einf\"uhrung in Operations Research}}.
\newblock Springer, Berlin.

\bibitem[Kreher \& Stinson, 1998]{Kreher:98}
Kreher, D., \& Stinson, D. (1998).
\newblock {\em {Combinatorial Algorithms: Generation, Enumeration and Search}}.
\newblock CRC Press.

\bibitem[Lu, 1996]{Lu:96}
Lu, G. (1996).
\newblock {\em {Communication and Computing for Distributed Multimedia
  Systems}}.
\newblock Artech House, Boston, London.

\bibitem[Mitchell, 1996]{Mitchell:96}
Mitchell, M. (1996).
\newblock {\em {An Introduction to Genetic Algorithms}}.
\newblock The MIT Press.

\bibitem[Nievergelt, 1977]{Nievergelt:77}
Nievergelt, J. (1977).
\newblock {\em {Combinatorial Algorithms}}.
\newblock Prentice Hall.

\bibitem[Pan, 1995]{Pan:95}
Pan, D. (1995).
\newblock {A Tutorial on MPEG/Audio Compression}.
\newblock {\em IEEE Multimedia}, 2(2),60--74.

\bibitem[Robertson, 1977]{Robertson:77}
Robertson, S.~E. (1977).
\newblock {The Probability Ranking Principle in IR}.
\newblock {\em Journal of Documentation}, 33(4),294--304.

\bibitem[Sch\"{a}uble, 1997]{Schauble:97}
Sch\"{a}uble, P. (1997).
\newblock {\em {Multimedia Information Retrieval---Content-Based Information
  Retrieval from Large Text and Audio Databases}}.
\newblock Kluwer Academic Publishers, Boston/London/Dordrecht.

\bibitem[Sutcliffe et~al., 1997]{Sutcliffe:97}
Sutcliffe, A., Hare, M., Doubleday, A., \& Ryan, M. (1997).
\newblock {\em {Intelligent Multimedia Information Retrieval}}, chapter
  {Empirical Studies in Multimedia Information Retrieval}, pp. 449--472.
\newblock AAAI Press.

\bibitem[Vidal, 1993]{Vidal:93}
Vidal, R. (1993).
\newblock {\em {Applied Simulated Annealing}}.
\newblock Springer.

\bibitem[Vorhees \& Harman, 1999]{Vorhees:99}
Vorhees, E., \& Harman, D. (1999).
\newblock {Overview of the Seventh Text REtrieval Conference (TREC-7)}.
\newblock In {\em {TREC-7 Proceedings}}.

\bibitem[Wechsler, 1998]{Wechsler:98a}
Wechsler, M. (1998).
\newblock {\em {Spoken Document Retrieval Based on Phoneme Recognition}}.
\newblock PhD thesis, ETH Zurich.
\newblock Diss. No. 12879.

\end{thebibliography}

\end{document}